\documentclass[12pt,twocolumn]{iopart}
\usepackage[usenames,dvipsnames]{color}
\usepackage{graphicx}
\usepackage[utf8]{inputenc}
\usepackage[T1]{fontenc}

\usepackage{ulem} % for striking out text
\begin{document}
\title{Directed transport   in coupled noisy Josephson junctions controlled via   ac signals}
\author{L. Machura, J. Spiechowicz and J. {\L}uczka}
\address{Institute of Physics, University of Silesia, Katowice, Poland}
\ead{jerzy.luczka@us.edu.pl}

\begin{abstract}
Transport properties of two coupled Josephson junctions  driven by
ac currents and thermal fluctuations  are studied with the purpose of determining 
dc voltage characteristics. It is a physical realization of directed transport induced by a non-biased zero averaged external signal.
 The ac current is applied either to (A)  only one junction as a biharmonic current  or (B)  is split into two simple harmonic components   and separately applied  to respective junctions.  We identify regimes
 where junctions can operate 
 with the same as well as opposite signs of voltages.  A general observation is that in the same parameters regimes,  the scenario (B) is more efficient in the sense that the induced dc voltages take  greater values. 
\end{abstract}

\submitto{\PS}
\maketitle

\section{Introduction}
Noisy transport in periodic arragements \cite{LucBar1995} is widely present 
in many nowadays branches of science --
in physics, biology, chemistry, economy and many others. On the physical ground,  the 
periodicity itself can be associated either with space degrees of freedom like in the crystals, 
optical lattices, systems of ring topologies  or with time-periodic drivings  like ac currents, magnetic 
or electric fields, rocking and pulsating forces to name but the few. It also can
be present in these both domains. Typical realizations can range from biophysics 
\cite{noiseinbio} with the description of   biomotors movement 
 on  asymmetric periodic microtubules \cite{biomotors} or  transport inside  ion
channels \cite{ionchannels}, to the present  experiments with optical lattices
\cite{Renzoni2005,Salger2012}, quantum mesorings \cite{mesorings} or Josepshon junctions
\cite{jj}. 

The Josephson effect is known for a half of the century \cite{Jos}. Since that time it
has been utilized for the definition of the voltage standard \cite{kautz} or for more 
practical devices as elements in high speed circuits \cite{jj} or even for the future 
applications in quantum computing devices \cite{qcjj}. Surprisingly, after 50 years of
intensive theoretical and experimental research, we are still  able to find new and 
uncommon phenomena even in a simple system of two weakly connected  superconductors. Recently,  the counterintuitive phenomenon  of  absolute negative
conductance (ANC) has been reported in the single  driven, resistively, and capacitively shunted
Josephson junction device subjected to both a time-periodic (ac) and a constant biasing (dc) current \cite{MacLuc2007}.  The ANC phenomenon  has been confirmed by the suitable
experiment with a Josephson junction setup \cite{NagSpe2008} and  very recently with ultracold atoms 
in  optical lattices \cite{Salger2012}. 
Other aspects of anomalous transport phenomena like 
 the occurrence of a negative differential conductance  and the emergence of a negative nonlinear conductance in the nonequilibrium response regime remote from zero dc bias have been studied in a series of  papers   \cite{ANM}. The  influence of the unbiased 
biharmonic ac current on a single junction has been considered in Refs. \cite{JJbi}. Recently the dynamics of 
the phase difference of coupled junctions has been addressed 
\cite{JanLuc2011,MacSpi2012}.

This work is organized as follows. In  \sref{sec2},  we present the model of two interacting 
junctions. Next,  in  \sref{sec3}, the numerical investigation of  transport properties
for two scenarios (A) and (B) of drivings applied to two coupled junctions is compared. The paper ends with summary and
conclusions in  section 4. 

\section{Model of driven interacting junctions}\label{sec2}
\begin{figure}[htbp]
    \centering
    \includegraphics[width=0.5\linewidth]{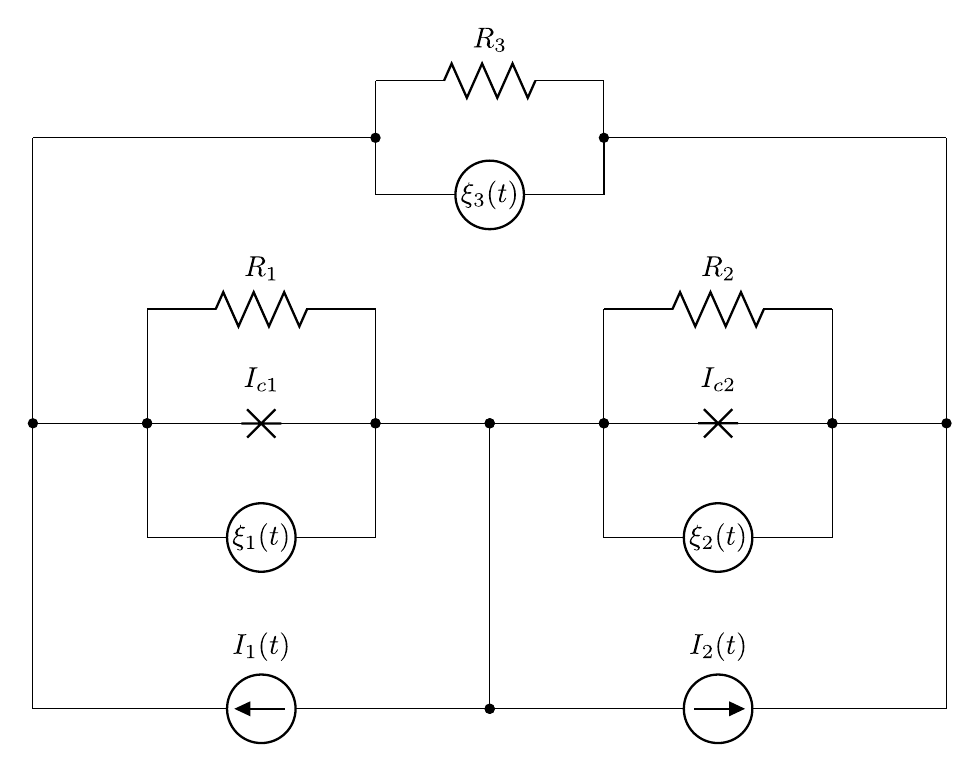}
    \caption{The system of two coupled Josephson junctions characterized by the critical Josephson supercurrents $(I_{c1}, I_{c2})$,
resistances $(R_{1}, R_{2})$, shunted by the  external resistance 
	$R_{3}$, influenced by  Johnson-Nyquist thermal noise sources 
$(\xi_{1}(t), \xi_{2}(t), \xi_{3}(t))$ and driven by the  external currents $(I_{1}(t), I_{2}(t))$.}
	\label{fig1}
\end{figure}

From a more general point of wiev,  we  explore the system consisting of two subunits (subsystems) interacting with each other.
The system is  driven out of its equilibrium state by an external force. As a particular realization
of this idea we  propose two resistively shunted Josephson junction devices 
characterized by the critical Josephson supercurrents $(I_{c1}, I_{c2})$,
resistances $(R_{1}, R_{2})$ and phases  $(\phi_{1}, \phi_{2})$ \cite{Ner80}. A schematic circuit 
representing the model is shown in \fref{fig1}. The system is externally shunted by the resistance 
$R_{3}$ and driven by two current sources $I_1(t)$ and $I_2(t)$ acting on the first and second junction, 
respectively. We also include into the model Johnson-Nyquist thermal noise sources 
$\xi_{1}(t), \xi_{2}(t)$ and $\xi_{3}(t)$ associated with the corresponding resistances 
$R_1, R_2$ and $R_3$ according to the fluctuation--dissipation theorem. 

The beauty of the standard Josephson theory lies in the simplicity of the model. In the  semiclassical regime, when 
the spatial dependence of characteristics can be neglected and photon-assisted tunnelling phenomena do not contribute,  the so-called Stewart-McCumber model \cite{stewart} holds true (for the extensive discussion on the validity of the model we refer the reader to \cite{kautz}). In this regime, one can use  the classical Kirchhoff current 
and voltage laws, and two Josephson relations to derive  two  evolution equations for the
phases $\phi_1=\phi_1(t)$ and $\phi_2=\phi_2(t)$. 
The dimensional version of the  equations is presented in Ref. \cite{MacSpi2012}. 
Here,  we recall their dimensionless form, namely, 
\numparts
\begin{eqnarray}\label{fi}
\dot{\phi}_1 &= I_1(\tau)- I_{c1} \sin \phi_1 
+\alpha [I_2(\tau)  -  I_{c2} \sin \phi_2] + \sqrt{D}\; \eta_1(\tau)\label{fi1},\\
\dot{\phi}_2 &= \alpha \beta[I_2(\tau) -  I_{c2} \sin \phi_2]
+ \alpha [ I_1(\tau) - I_{c1} \sin \phi_1] + \sqrt{\alpha \beta D} \;\eta_2(\tau)\label{fi2},   
\end{eqnarray}
\endnumparts
where $\phi_i=\phi_i(\tau)$ for $i=1,2$ and the dot denotes a derivative with the respect to the dimensionless time $\tau$ expressed by the dimensional time $t$  as
\begin{equation}
\label{time}
\tau = \frac{2eV_0}{\hbar} t, \quad 
V_0 =  I_c \frac{R_1(R_2+R_3)}{R_1+R_2+R_3}, \quad 
I_c=\frac{I_{c1}+I_{c2}}{2}. 
\end{equation}
The parameters 
\begin{eqnarray}
  \label{alfa}
  \alpha = \frac{R_2}{R_2 + R_3}  \in [0, 1], \quad \beta = 1 + \frac{R_3}{R_1}, \quad D = \frac{4ek_{B}T}{\hbar I_{c}}. 
\end{eqnarray}
All dimensionless currents $I_1(\tau), I_2(\tau), I_{c1}$ and $I_{c2}$ are in units of $I_c$, e.g.   $I_{c1}  \to I_{c1}/I_c$. 
Thermal equilibrium noise sources related to the resistances $R_1, R_2, R_3$ are modelled here 
by the independent $\delta$--correlated zero-mean Gaussian white noises $\xi_{i}(t)\ (i = 1,2,3)$, i.e., 
$\left< \xi_{i}(t)\xi_{j}(s)\right> = \delta_{ij}\delta(t - s)$ for $i, j \in \{1, 2, 3\}$.
The straightforward assumption of identical temperature $T$ felt by all parts
of the set--up allows for the reduction of the number of original noises $\xi_1, \xi_2, \xi_3$ (see \fref{fig1}),  to their
linear combination $\eta_1$ and $\eta_2$ in the equations (\ref{fi1}) and (\ref{fi2}).  

The reader would find easier to understand this scenario
within a pure mechanical picture. The dynamics of the phase difference can be mapped
onto the motion of the Brownian particle. In this mechanical analog the correspondence
between position $x_{1}$ of the first particle with the phase difference
$\phi_{1}$ of the first junction  can be settled and the position $x_2$
of the second particle can mimic the phase difference  $\phi_2$ of the second 
junction. If we imagine two interacting particles moving along the  
periodic structure than the most significant quantifiers describing their transport properties  would be 
the average velocities of the first $v_1 = \langle\dot{\phi_{1}}\rangle$ and second 
$v_2 = \langle\dot{\phi_{2}}\rangle$ particle, respectively. In terms of the Josephson junction system it corresponds 
to the dimensionless long-time averaged voltages $v_1 = \langle\dot{\phi_{1}}\rangle$ and 
$v_2 = \langle\dot{\phi_{2}}\rangle$ across the first and second junctions, respectively 
(from the Josephson relation, the dimensional voltage $V=(\hbar/2e)d\phi/dt$ and therefore  
$d\phi/d\tau = V/V_0$). The junction resistances (or conductance) translates then into the particles mobility.
Moreover, the phase space of the deterministic system is three-dimensional 
$\{\phi_{1}(\tau), \phi_{2}(\tau), \omega \tau \}$ and therefore give rise to possible
chaotic evolution which is  the key feature for anomalous transport 
\cite{MacLuc2007,ANM,reim}.

%%%%%%%%%%%%%%%%%%%%%%%%%%%%%%%%%%%%%%%%%%%%%%%%%%%%%%%%%%%%%%%%%%%%%%%%%%5

\subsection{Identical junctions}\label{subsec21}

Without loss of generality, we can reduce a number of parameters assuming that two junctions are \textit{identical} with 
 $R_{1} = R_{2}$ and $I_{c1} = I_{c2} \equiv 1$. In such a case $\alpha \beta = 1$ and 
equations \eref{fi1} and \eref{fi2} take  symmetric form
\numparts
\begin{eqnarray}
\label{idfi1}
\dot{\phi}_1 &=I_1(\tau)-\sin \phi_1+\alpha [I_2(\tau)-\sin \phi_2]+\sqrt{D}\;\eta_1(\tau),\\
\label{idfi2}
\dot{\phi}_2 &=I_2(\tau)-\sin \phi_2+\alpha [I_1(\tau)-\sin\phi_1]+\sqrt{D}\;\eta_2(\tau).     
\end{eqnarray}
\endnumparts
The  parameter $\alpha = R_2/(R_2 + R_3) \in [0,1]$  
plays the role of coupling strength between the junctions and can be tuned by the 
variation of the external resistance $R_3$. When $\alpha =0$  the set of equations (4) decouple into two independent equations. It can be realized taking $R_3 \to \infty$.  The opposite situation with  two fully coupled junctions can be worked out by designating $R_3=0$.
The noise strength $D$ can be tuned by temperature. The currents  $I_1(\tau)$ and $I_2(\tau)$ are energy sources pumped into the system and 
can be applied to one or to both junctions.

%%%%%%%%%%%%%%%%%%%%%%%%%%%%%%%%%%%%%%%%%%%%%%%%%%%%%%%%%%%%%%%%%%%%%%%

\subsection{External current driving}\label{subsec22}

The trivial way to induce the dc voltage across both 
junctions is to apply the dc current to both junctions separately (in the mechanical analog, it corresponds to the static force). It seems that we
can also do it by applying the dc current to one junction only but have to make sure that coupling is  strong enough to
call out the response on the other junction too. This, however, seems to be rather uninteresting
and a well known solution. What if we abandon simple intuitive possibilities? We can exploit the well known
ratchet effect \cite{bmbible} and induce the non--zero dc voltage  by applying a zero-mean external current. We consider two scenarions. In the first scenario (A), the 
 ac driving to applied to only one of the junctions \cite{JJbi,MacSpi2012}, namely,  
\begin{equation}
\label{drv1}
I_1(\tau) = a_1 \cos(\omega \tau) + a_2 \cos(k \omega \tau + \theta),\qquad I_2(\tau) = 0. 
\end{equation}
where $\theta$ is the relative phase between the driving currents and $k$ is a real number. 

In the second scenario (B), the external current is  split into two simple harmonic components applied to two respective junctions, namely, 
\begin{equation}
\label{drv2}
I_1(\tau) = a_1 \cos(\omega \tau),\qquad I_2(\tau) = a_2 \cos(k \omega \tau + \theta).
\end{equation}
 We know that the symmetric driving cannot 
itself induce the non--zero dc voltage.  However,  we expect that the coupling between junctions  would 
have to play the crucial role in the dynamics of the total system and a non-zero dc voltage could be generated for  $\alpha > 0$. 
We ask which of two scenarios (5) or  (6) is more efficient in the sense that the induced dc voltages have  greater amplitudes.    
In the method (5) we have the  possibility to induce the non--zero dc voltage just by the ratchet effect, cf. the detailed discussion in Ref.  \cite{MacSpi2012}. In this case,  even for $\alpha =0$  we still can find  
non--zero dc voltage across the first junction. In the scenario (6) the separated symmetric ac currents 
cannot alone induce non--zero voltage in the decoupled junctions. Setting the parameter $\alpha \ne 0$ 
we effectively incorporate the ratchet effect and in turn create the prospect of dc transport in the system.

%%%%%%%%%%%%%%%%%%%%%%%%%%%%%%%%%%%%%%%%%%%%%%%%%%%%%%%%%%%%%%%%%%%%%%

\section{Dc voltage characteristics}\label{sec3}

Stochastic differential equations (4) cannot  be handled by known analytical methods. For this reason we have carried out extensive numerical simulations. 
We have used the $2^{nd}$ order Stochastic Runge-Kutta algorithm with the time step of about
$10^{-3} \cdot (2\pi/\omega)$. The initial phases $\phi_{1}(0)$ and $\phi_{2}(0)$ have been 
randomly chosen from the interval $[0, 2\pi]$. Averaging was performed over $10^{3} - 10^{6}$ 
different realizations and over one period of the external driving $2\pi/\omega$. Numerical 
simulations have been carried out using CUDA environment on desktop computing processor NVIDIA 
GeForce GTX 285. This gave us possibility to speed up the numerical calculations up to few 
hundreds times more than on typical modern CPUs \cite{cuda}. 
Below, we present results for a fixed  frequency multiplier $k = 2$.   This will reflect
the typical biharmonic driving studied previously for Hamiltonian systems \cite{flach}, 
systems in the overdamped regime \cite{Borromeo2005a,flachepl} and for the moderate damping 
\cite{Breymayer1984,MacLuc2010}. If the given parameter is not addressed directly in the plot we will
keep the constant values as follows: the noise strength (or 
equivalently the dimensionless temperature) $D=0.001$, the frequency of the ac 
driving $\omega = 0.03944$, the coupling strength $\alpha=0.56$, the relative
phase $\theta = \pi/2$ and the amplitudes $a_1 = a_2 = 1$.

\begin{figure}[htbf]
    \centering
	\includegraphics[width=0.49\linewidth]{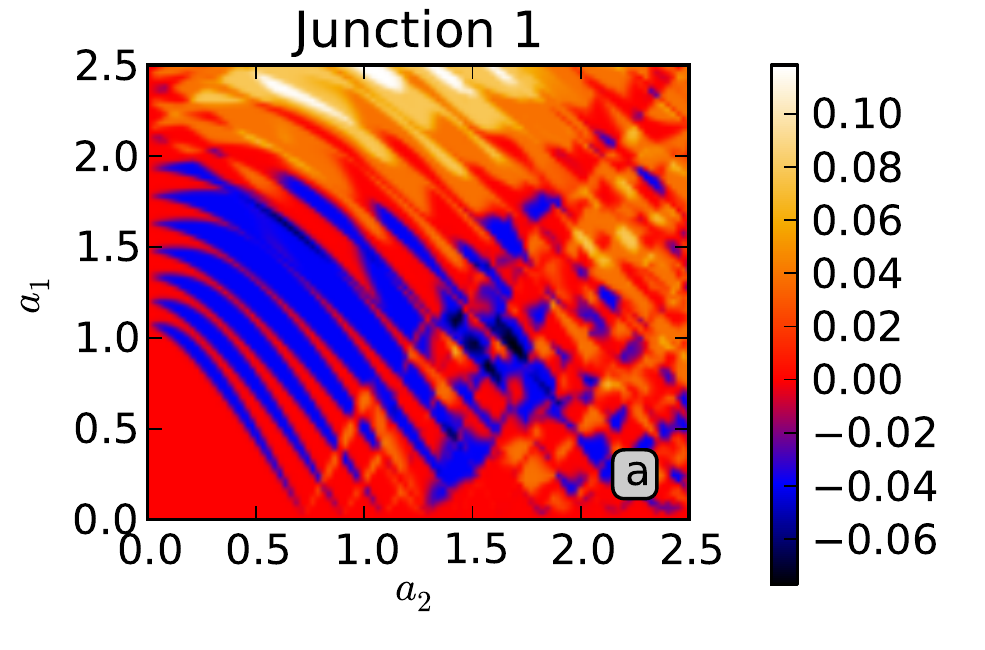}
	\includegraphics[width=0.49\linewidth]{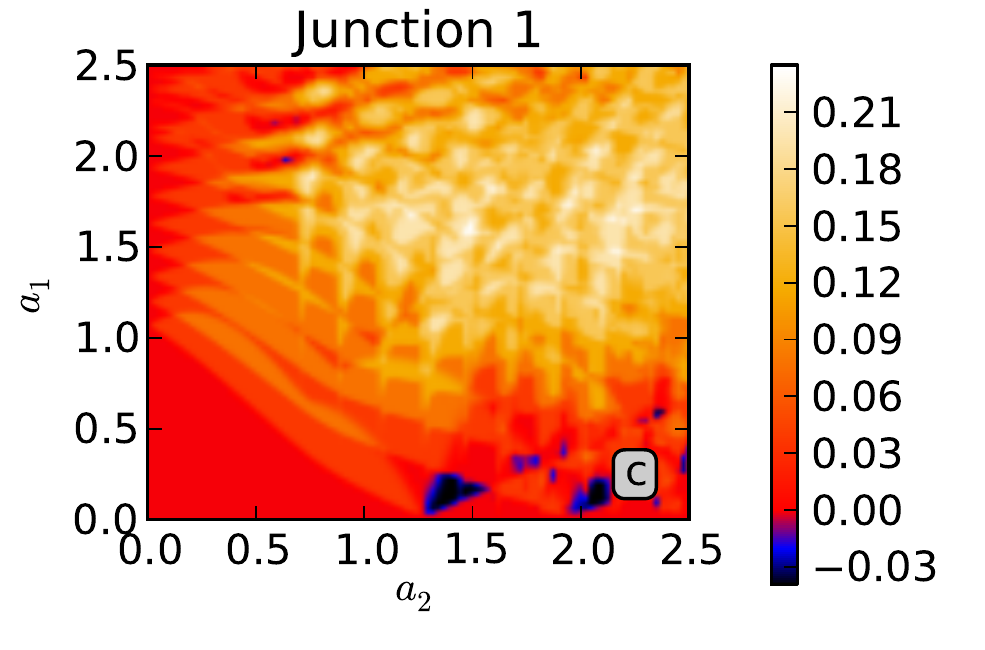}\\
	\includegraphics[width=0.49\linewidth]{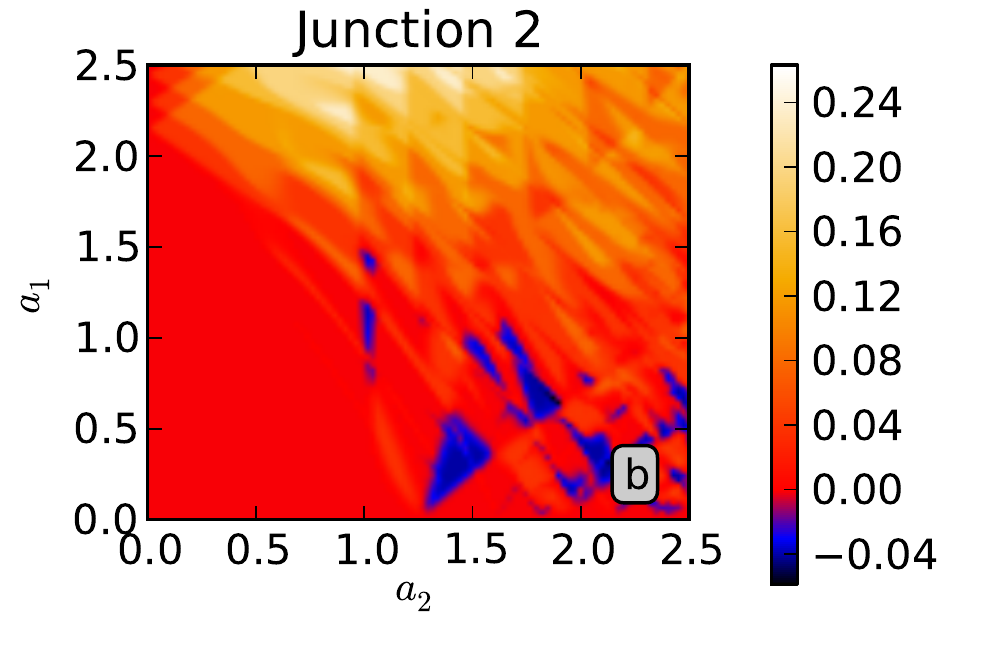}
	\includegraphics[width=0.49\linewidth]{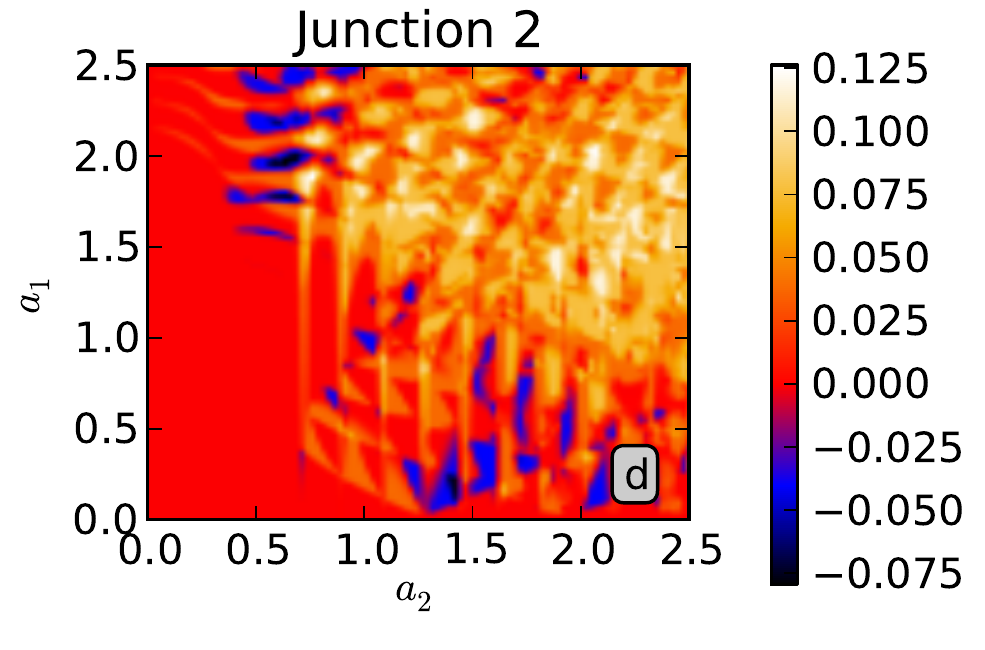}
	\caption{(color online) The stationary averaged dc voltages $v_1$ and $v_2$ across 
	the first and second junction. 
  The dependence on the external ac current amplitudes $a_1$ and $a_2$ are presented 
  in  panels (a) and (b) for the driving (5) acting  on the first junction only  and in  panels (c) and (d) for the driving (6) split  between two 
  junctions. 
  Other parameters read: the dimensionless temperature $D=0.001$, the 
  frequency $\omega = 0.03944$, coupling strength $\alpha=0.56$, the relative
  phase $\theta = \pi/2$ and the frequency multiplier $k = 2$.
  }
	\label{fig2}
\end{figure}
In the long time limit, the averaged voltages $\langle {\dot \phi_i(\tau)} \rangle$  
can be presented in the form of a series of all possible harmonics, namely,  
\begin{eqnarray}
\label{asym}
\lim_{\tau\to\infty} \langle {\dot \phi_i(\tau)} \rangle =
v_i + \sum_{n=1}^{\infty} v_i(n \omega \tau), \quad  i = 1, 2, 
\end{eqnarray}
where $v_i$ is a dc (time-independent) component and $v_i(n \omega t)$ are 
time-periodic functions of zero average over a basic period. For high frequency $\omega$ 
(i.e. fast alternating currents)  the averaged dc voltages are zero: very fast positive and negative changes 
of the driving current cannot induce the dc voltage and only multi-harmonic components of the voltages can survive.
In addition, if both amplitudes $a_1, a_2$ are smaller than the critical supercurrents, from the structure of the model (4)  
 it follows that the net voltage will be zero or very close to zero. 
%Transport sets in if at least one amplitude is of the order of unity. 

\begin{figure}[htbf]
	\centering
	\includegraphics[width=0.49\linewidth]{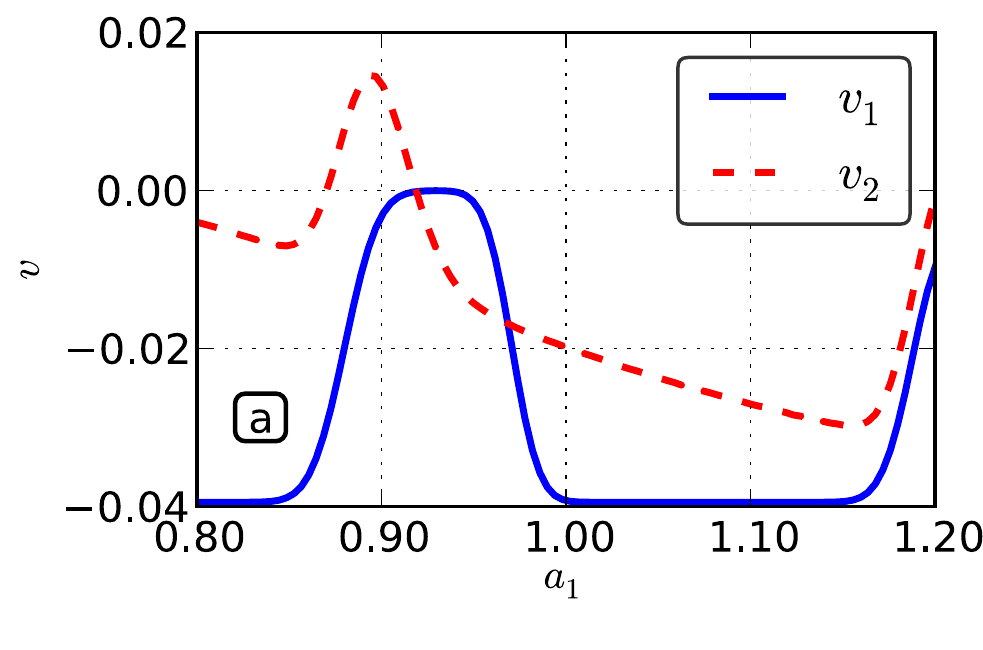}
	\includegraphics[width=0.49\linewidth]{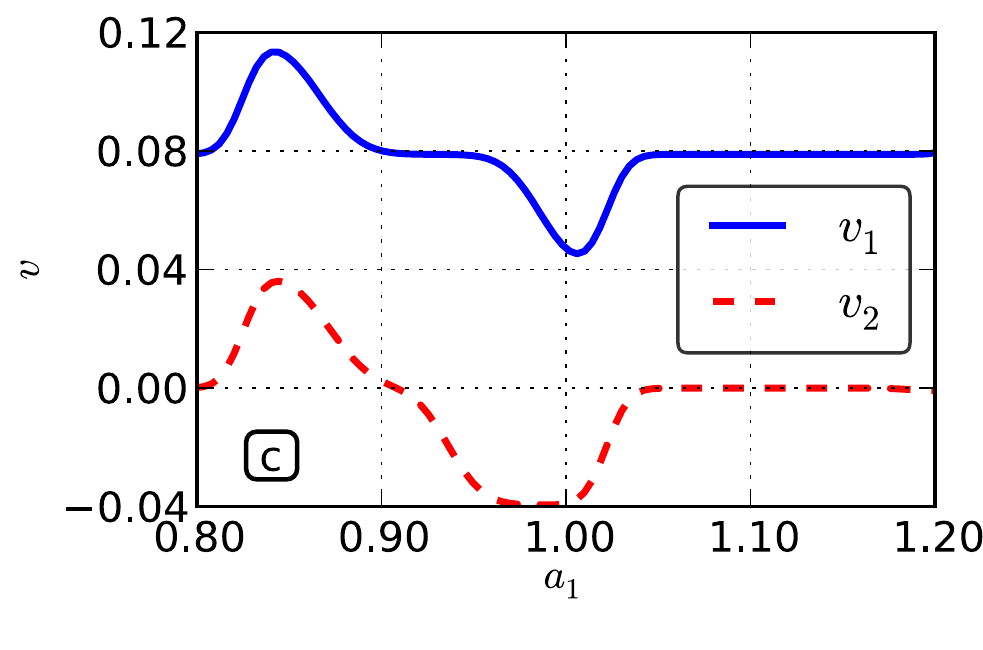}\\
	\includegraphics[width=0.49\linewidth]{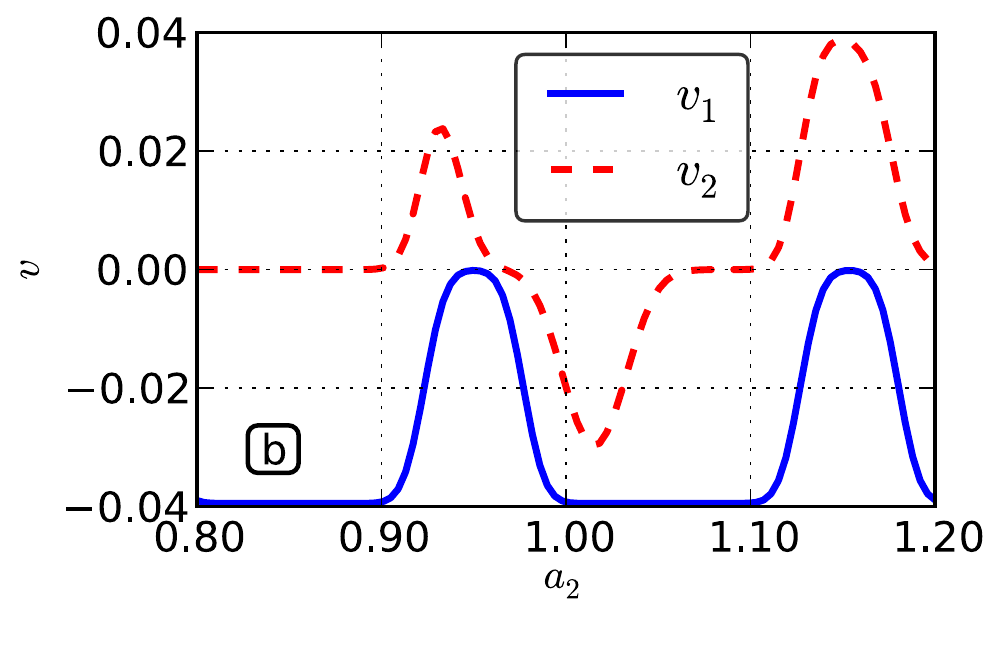}
	\includegraphics[width=0.49\linewidth]{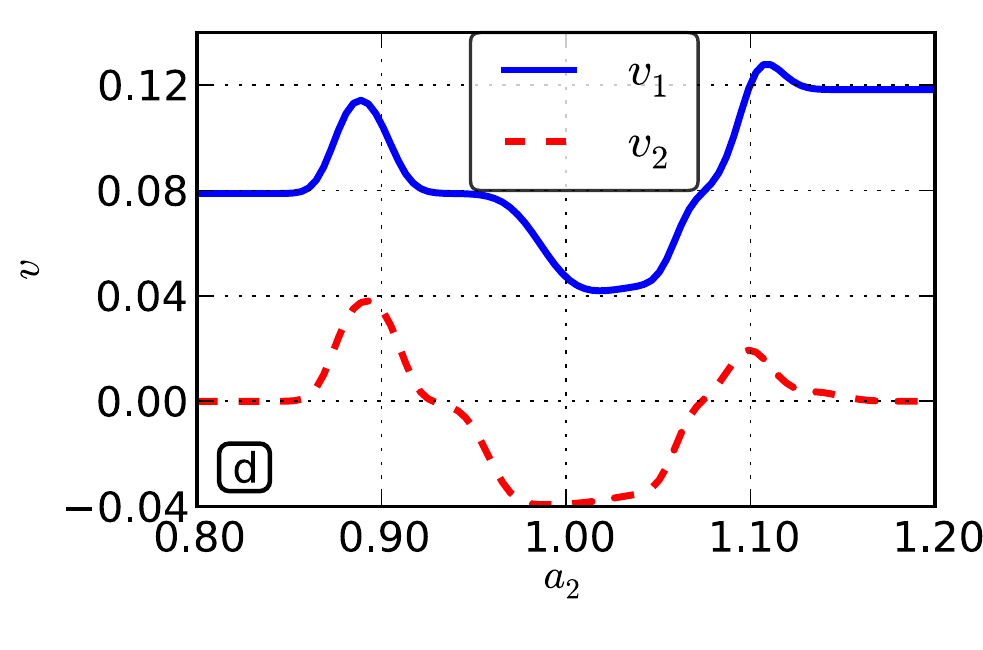}
	\caption{(color online) The stationary averaged dc voltages $v_1$ and $v_2$ across 
	the first (\textbf{{\color{blue}\full}} blue line) and second 
  (\textbf{{\color{red}\dashed}} red line) junction, respectively. 
  The dependence on the external ac current amplitudes $a_1$ and $a_2$ are presented for the driving (5)
  in  panels (a) and (b)  and  for the driving (6) in  panels (c) and (d). 
  Other parameters read: the dimensionless temperature $D=0.001$, the 
  frequency $\omega = 0.03944$, coupling strength $\alpha=0.56$, the relative
  phase $\theta = \pi/2$ and the frequency multiplier $k = 2$ and the amplitudes $a_1=a_2=1$.
  }
	\label{fig3}
\end{figure}
In    \fref{fig2}  the long-time  averaged 
dc voltages across the first ($v_1$) and  second ($v_2$) junctions are shown
in the amplitudes parameter plane $\{a_1, a_2\}$. There is clearly zero average voltage for
small values of both amplitudes. However, for larger amplitudes in  both scenarios (5) and (6) we can recognise 
four operating regimes where:
\begin{itemize}
\item[(i)] $v_1 > 0$ and $v_2 > 0$,
\item[(ii)] $v_1 < 0$ and $v_2 < 0$,
\item[(iii)] $v_1 < 0$ and $v_2 > 0$,
\item[(iv)] $v_1 > 0$ and $v_2 < 0$,
\end{itemize}
 Of course, the quantitative picture is different. One
can easily see that for the case (5),  transport properties of the first (driven) junction  have  more complicated strips-like structure with larger  area
of negative voltage. For the second (non--driven) junction, the dc voltage $v_2$ can be twice - three times greater than $v_1$.   On the other hand, in the case (6), the regimes of negative voltage are smaller. We emphasize that such complicated regimes of islands and tongues
of negative and positive  dc voltages  are not just rare occurrences: they can be verified with
numerically arbitrarily-high-accuracy calculations and over
extended intervals in the parameter space.

\begin{figure}[htbf]
	\centering
	\includegraphics[width=0.49\linewidth]{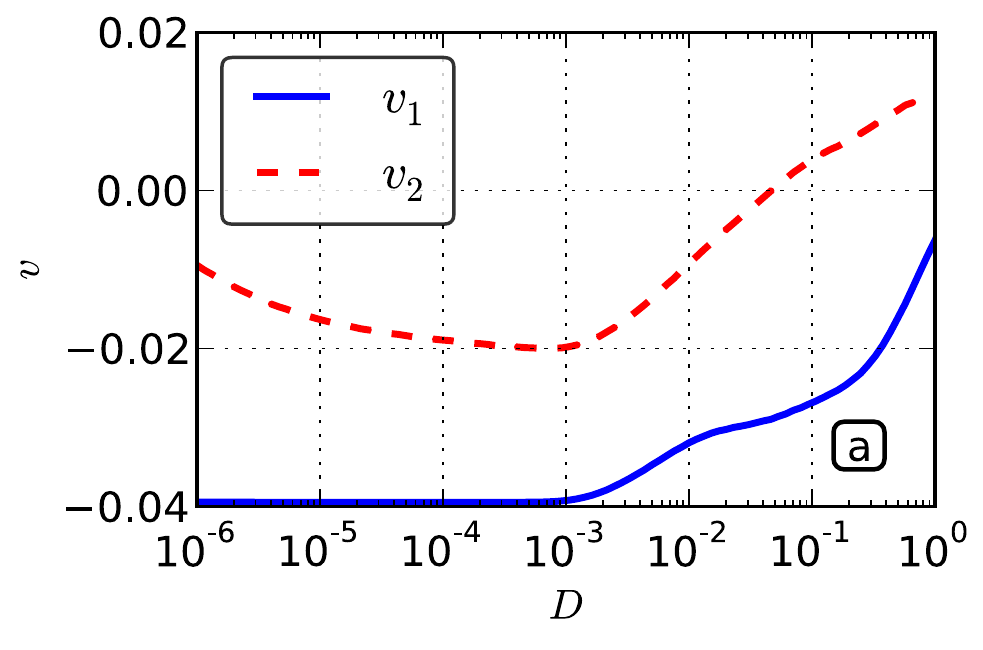}
	\includegraphics[width=0.49\linewidth]{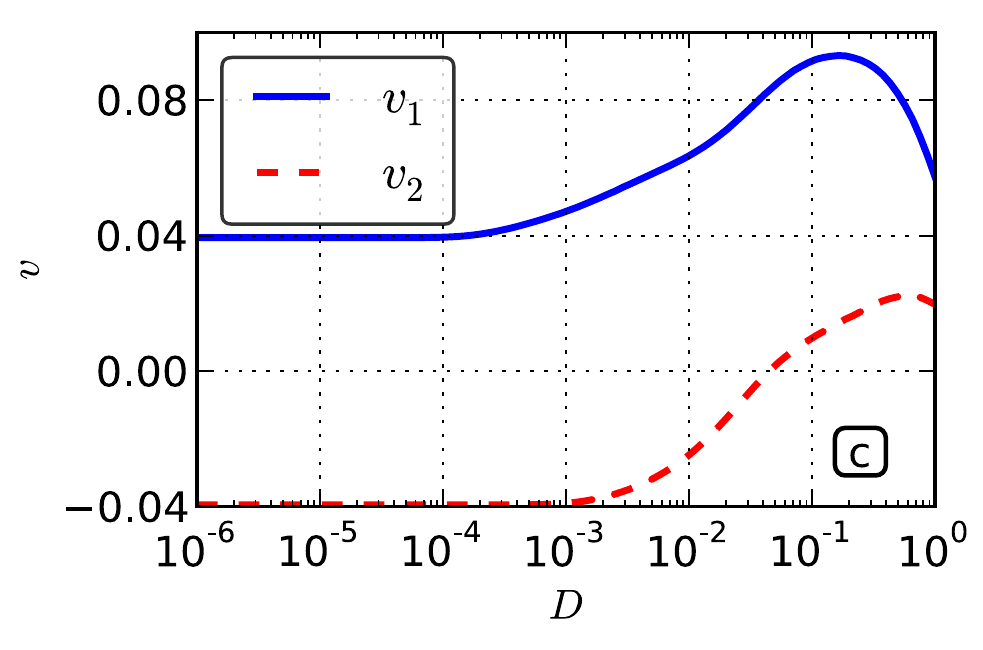}\\
	\includegraphics[width=0.49\linewidth]{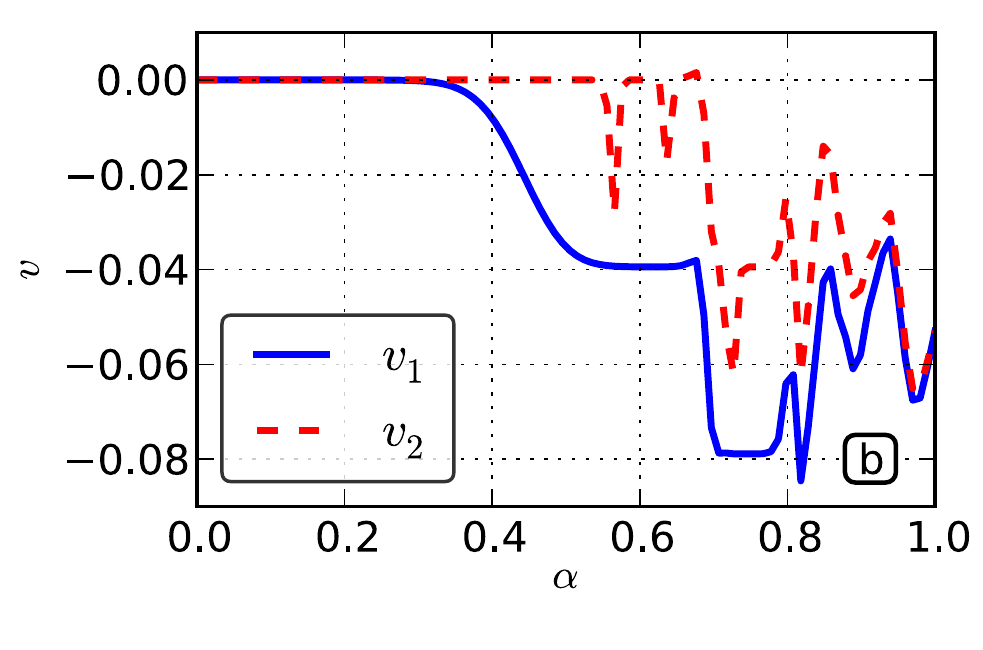}
	\includegraphics[width=0.49\linewidth]{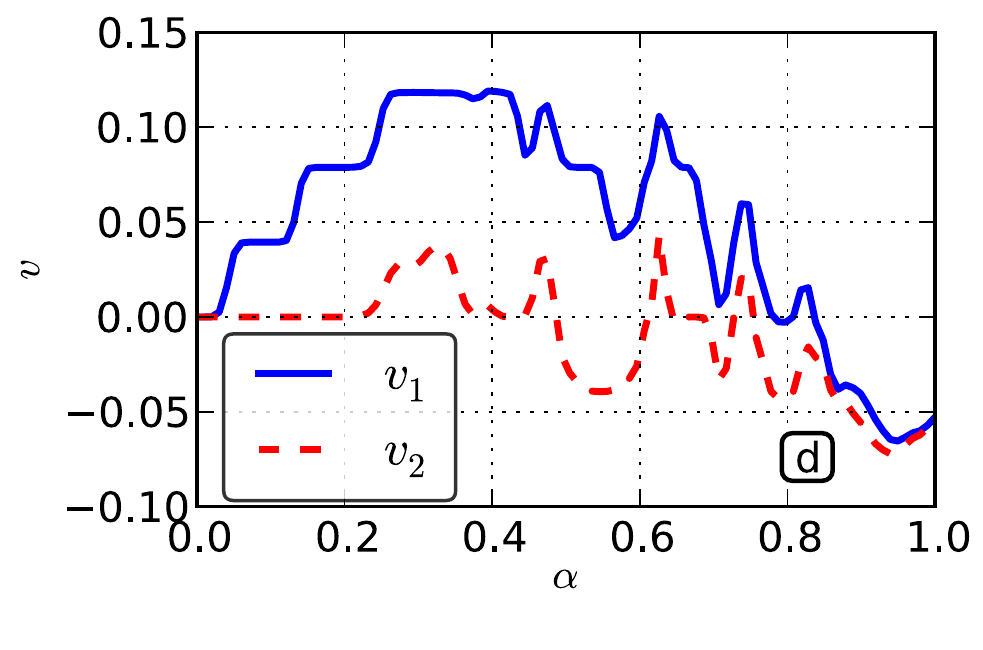}
	\caption{(color online) The stationary averaged dc voltages $v_1$ and $v_2$ across 
	the first (\textbf{{\color{blue}\full}} blue line) and second 
  (\textbf{{\color{red}\dashed}} red line) junction, respectively. 
  The dependence on 
  the noise strength (or dimensionless temperature) $D$ (upper  panels) and the 
  coupling strength $\alpha$ (bottom panels) are presented for the driving (5) 
  in  panels (a) and (b),  and for the driving (6) in  panels (c) and (d). 
  Other parameters if not addressed directly in the plots read: the dimensionless temperature $D=0.001$, 
  the frequency $\omega = 0.03944$, coupling strength $\alpha=0.56$, the relative
  phase $\theta = \pi/2$, the current amplitudes $a_1 = a_2 = 1$ and the frequency multiplier $k = 2$.
  }
	\label{fig4}
\end{figure}
In \fref{fig3}, we have selected the region where
for the scenario (5)  the dc voltage on the first (driven) junction  stays negative or zero (more precisely, so  small to be negligible)  
throughout the presented range of both amplitudes. The second (non-driven) junction 
shows all possible working states with the negative, positive and zero dc voltage 
(see left panels  of \fref{fig3}). In the same region but for the second scenario (6),  the first junction 
driven by the current  $\cos\omega\tau$   stays positive  for  
all presented values of both current amplitudes $a_1$ and $a_2$, while the junction driven by the current  $\cos(2\omega \tau + \pi/2)$ can assume  positive and negative values. 
What is also striking is that in the case (6) the voltage characteristics of both junctions change with some synchrony: the voltages increase or decrease when one of the current amplitude varies.   
 This feature, in turn, is not found in the scheme (5).

As the next point of analysis, we ask about the 
role of  thermal fluctuations.  
It  is presented in upper panels of 
\fref{fig4}. In this regime we can note the voltage reversal across the second junction:  the voltage $v_2$  can change its sign from negative  to positive  values when temperature is increased. On the other hand, the voltage $v_1$ is always negative for the scenario (5) and is always positive  for the scenario (6). For high temperature, both voltages tend to zero.  Next, we  address the issue of whether, and to which extent,
the coupling strength $\alpha$ can influence voltage properties. The results are depicted in  bottom panels of \fref{fig4}. The first note is non-monotonic and irregular dependence of both volatges on $\alpha$ with several minima and maxima. In the scenario (6), a step-like dependence of the voltage across the first junction is observed for small values of the coupling $\alpha$.

\section{Summary}

With this study we numerically analyzed the role of ac current  drivings  on   transport properties of the resistively shunted two  coupled Josephson junctions.   We identified a rich variety of the dc voltage  characteristics in the  parameter space where
 transport can be experimentally monitored. We have  detected  regions displaying  positive and negative dc voltages, which   form complicated structures in the  parameter space. 
We have mainly concentrated the analysis on impact of
selected regimes in the parameter space on voltage  properties. 
 Other regimes of parameters also modify voltage characteristics but here  we do not present all varieties.   
 A general observation is that in the same parameters regimes,  the biharmonic ac driving applied only to  one junction results is a smaller dc voltage than in the case  when the ac current  is split into two simple harmonics and each applied to  respective junctions. 
 
\ack
The work supported in part by the grant N202 052940 and 
the ESF Program "Exploring the Physics of Small Devices".

%%%%%%%%%%%%%%%%%%%%%%%%%%%%%%%%%%%%%%%%%%%%

\end{document}